\documentclass{aa}
\usepackage{natbib}
\usepackage{amssymb}
\usepackage{amsmath}
\usepackage{graphicx}
\usepackage{xspace}
\def\revised#1{{#1}}

\def\aap{A\& A}

\def\araa{AnnRevA\& A}
\def\mnras{MNRAS}
\def\apj{ApJ}

\def\apjl{ApJL}

\def\nat{Nature}

\def\in{\mathrm{in}}
\def\rim{\mathrm{warm}}
\def\out{\mathrm{out}}
\def\ssd{\mathrm{ssd}}
\def\isaf{\mathrm{isaf}}
\def\warm{\mathrm{warm}}

\def\xx{\mathrm{x}}

\def\rssdin{\ensuremath{R_\mathrm{ssd}}\xspace}
\def\rwarmin{\ensuremath{R_\mathrm{warm}}\xspace}

\def\sigssd{\ensuremath{\Sigma_\mathrm{ssd}}\xspace}
\def\rhoisaf{\ensuremath{\rho_\mathrm{isaf}}\xspace}
\def\tisaf{\ensuremath{T_\mathrm{isaf}}\xspace}

\def\evapssdwarm{\ensuremath{\psi_{\mathrm{ssd}\rightarrow\mathrm{warm}}}\xspace}
\def\evapwarmadaf{\ensuremath{\psi_{\mathrm{warm}\rightarrow\mathrm{isaf}}}\xspace}
\def\fion{\ensuremath{\psi_{\mathrm{isaf}\rightarrow\mathrm{ssd}}}\xspace}
\def\ISAF{ISAF}
\def\comma{\,,}
\def\fullstop{\,.}
\def\thttle{Evaporation of ion-irradiated disks}
\begin{document}
\title{\thttle}
\author{C.P.~Dullemond and H.C.~Spruit}
\authorrunning{Dullemond \& Spruit}
\titlerunning{\thttle} 
\institute{Max-Planck-Institut f\"ur Astrophysik, Karl--Schwarzschildstrasse 1, D--85748 Garching, Germany; e--mail:
dullemon@mpia.de}
\date{DRAFT, \today}

\abstract{We calculate the evaporation of a cool accretion disk around a
black hole due to the ion-bombardment by an ion supported accretion flow
(here ISAF, or optically thin ADAF). As first suggested by Spruit \& Deufel
(\citeyear{spruitdeufel:2002}), this evaporation takes place in two stages:
ion bombardment of the cool disk (Shakura-Sunyaev disk: SSD) produces an
intermediate-temperature layer on top of the disk (`warm layer') which
constitutes an independent accretion flow on both sides of the SSD. As this
warm material accretes inward of the inner radius of the SSD, it becomes
thermally unstable by lack of cooling of photons, and evaporates into the
\ISAF{}, thereby feeding the latter. Angular momentum conservation forces a
certain fraction of the \ISAF{} material to move outward, where it can
bombard the SSD with its hot ions. The flow geometry is derived by computing
stationary solutions of the continuity- and angular momentum equations for
the three components (\ISAF{}, warm flow and SSD). The overall radiative
output is dominated by hard X-rays. They are produced mostly from the warm
component, rather than the \ISAF. The expected time dependence and stability
of the flow, not computed here, is discussed briefly.  }

\maketitle


\section{Introduction}
The X-ray spectra of galactic black holes occur in two different `states': a
soft state dominated by a soft multi-color blackbody component peaking
around 1 keV, and a hard state in which the energy output peaks at 100-200
keV. The soft state can be explained well with a standard optically thick
geometrically thin accretion disk of the kind of Shakura \& Sunyaev
(\citeyear{shaksuny:1973}). But what produces the emission in the hard state
is still under debate. Some argue that it represents non-thermal synchrotron
radiation from a jet (Markoff et
al.\citeyear{markofffalckefender:2001}). Others invoke magnetic activity on
the surface of the disk, creating a `corona' on top of the disk (Galeev et
al.~\citeyear{galeevrosvai:1979}; Haardt et
al.~\citeyear{haardtmarghis:1994}; Stern et al.~\citeyear{sternpout:1995};
Merloni \& Fabian \citeyear{merlonifabian:2001}).  A commonly cited
explanation for the hard X-ray spectra of black holes is the comptonization
of soft photons in of a hot optically thin virialized plasma flow. The
concept of such a hot `ion torus' around black holes has already been
proposed a long time ago (Shapiro, Lightman \& Eardley \citeyear{sle:1976};
Ichimaru \citeyear{ichimaru:1977}; Rees et al.~\citeyear{reesbegblph:1982}),
but optically thin plasmas in the 100 keV range are notoriously unstable
(Lightman \& Eardley \citeyear{lighteard:1974}; Sunyaev \& Shakura
\citeyear{sunyaevshakura:1975}; Piran \citeyear{piran:1978}; Chen et
al.~\citeyear{chenabrlas:1995}). In the 90s it was recognized that the
advection of thermal energy in an inefficiently radiating hot optically thin
plasma flow can have a stabilizing effect on the flow (Narayan \& Yi
\citeyear{narayanyi:1995}). Instead of converting potential energy into
radiation, such an ``Advection Dominated Accretion Flow'' (ADAF, here
called ``Ion Supported Accretion Flow'': ISAF) stores
the released potential energy into thermal energy of the ions and simply
advects it down through the event horizon. A crucial element in the theory
of such \ISAF{}s is the thermal decoupling of ions and electrons. If the
temperature of the electrons would be equal to the (almost virial)
temperature of the ions, the flow would cool dramatically and collapse. The
weakness of the Coulomb coupling between the eletrons and the ions allows
the electrons to have a much cooler temperature (of the order of $2\times
10^9$K) than the ions (which can have temperatures reaching $10^{11}$K), and
therefore to radiate only very limitedly, allowing the thermal energy to be
advected into the black hole and the hot flow to be stable. Despite the
relative inefficiency of \ISAF{}s, they {\em do} emit at least {\em some}
Bremsstrahlung and thermal synchrotron radiation, and they comptonize soft
photons to high energies, which could explain the hard X-ray components
observed in the spectra of black holes.

While this scenario seems appealing, there are a number of open issues. One of
these is the existence of `intermediate' states in which both components are
present {\em simultaneously}, suggesting that the hot \ISAF{} and the cool SSD
somehow coexist. Based on simple mean-free-path arguments one can show that if
these flows exist at the same distance from the black hole, the ions 
of the \ISAF{} (with energies of 1 -- 100 MeV) are quickly lost by penetrating
into the SSD, causing the \ISAF{} to condense. Another issue is that \ISAF{}s
can only exist up to a certain distance from the black hole,
because the two-temperature plasma state ceases to exist at low
temperatures, (e.g.~Igumenshchev et al.~\citeyear{igumabrnov:1997}; Czerny
et al.~\citeyear{czernyroz:2000}; Narayan \& Yi \citeyear{narayanyi:1995}).
Therefore their existence means that they must have been produced in situ,
probably by some evaporation mechanism acting on the cool Shakura-Sunyaev
disk (hereafter `SSD'): the \ISAF{} being the `vapor' from the SSD. A
well-studied evaporation mechanism is `coronal' evaporation (which
determines the chromosphere-corona transition in the solar atmosphere). It
is governed by the conduction of heat by the electrons (Meyer \&
Meyer-Hofmeister \citeyear{meyermeyhof:1994}). This has been shown to work
well for models of Cataclysmic Variables. In the two-temperature regime,
this process fails because the energy is not carried by the electrons any
more.  For black hole accretion the mechanism therefore only works well at
large radii ($\gtrsim 10^3\,R_{s}$). Yet there are strong indications that
some evaporation mechanism acts on the SSD at much smaller radii: from
analysis of X-ray spectra of the soft component it has been inferred that
the inner radius of the SSD regularly moves outward from the last stable
orbit to distances of a few tens of Schwarzschild radii, in particular
during the transition from a `high-soft' state to a `low-hard' state
(Zdziarski et al.~\citeyear{zdziarskilubsmith:1999}; Gilfanov et
al.~\citeyear{gilfchurrev:2000}; Churazov et
al.~\citeyear{churgilfrev:2001}; Revnivtsev et
al.\citeyear{revgilfchur:2001}).

The simultaneous presence of a hot comptonizing plasma close to the black
hole and an SSD with varying inner radius therefore poses a problem: what is
the mechanism that evaporates the SSD at close distances from the event
horizon, thereby producing the hot \ISAF{}? Some time ago Spruit \&
Deufel (\citeyear{spruitdeufel:2002}) proposed that a possible mechanism
could be in fact the injection of \ISAF{} ions into the SSD. Their
calculations show that such an ion bombardment produces an
intermediate-temperature (about 60 keV) layer of optical depth $\tau\sim1$
on top of the SSD surface (Deufel, Dullemond \&
Spruit \citeyear{deufeldulspruit:2002}).

The temperature of the warm layer is kept stably in this intermediate range
(which would be thermally unstable for an isolated optically thin plasma) by
the balance between heating and Compton cooling by the soft photon flux from
the cool disk. This is similar to the energy balance in the Haardt-Maraschi
model (1994), but the present model is more predictive, since both the
temperature and the optical depth of the layer are determined by the physics
of ion penetration.

Such a `warm layer', being of much lower density and much higher temperature
than the SSD (but still much lower than the virial temperature), transports
its own angular momentum and forms an accretion flow on top of the disk. Now
suppose the inner radius of the SSD ($\rssdin$) is significantly far outside
the last stable orbit. The radial flow velocity in the warm layer is orders
of magnitude higher than the flow velocity in the SSD, and therefore this
warm material quickly accretes inward of $\rssdin$. In this region there is
no SSD underneath the warm layer to provide soft photons for compton
cooling, and therefore the temperature of the warm accretion flow inward of
$\rssdin$ increases (to roughly 300 keV, Deufel et al.
\citeyear{deufeldulspruit:2002}). Due to the reduced soft photon flux, this
region is now also thermally unstable, and starts evaporating into the
\ISAF{} (Spruit \& Deufel 2002)

\begin{figure}
\centerline{
\includegraphics[width=9cm]{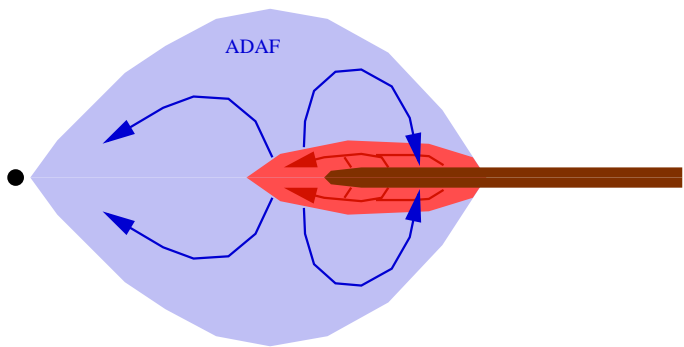}}
\caption{\label{fig-geometry}The geometry of the three-layered accretion
flow, in which evaporation of the cool disk takes place in two stages.
}
\end{figure}

This process produces the mass flux that sustains the \ISAF{}. Part of the
\ISAF{} material accretes inward, and part moves outward due to angular
momentum conservation (as in any viscous disk, e.g.\ Pringle
\citeyear{pringle:1981}; Liu et al.~\citeyear{liumeymey:1997}; Turolla \&
Dullemond \citeyear{turdul:2000}). The outward moving \ISAF{} material
spreads over the SSD. Over a short distance it dumps its ions onto the SSD,
thus producing the warm layer and closing the cycle. This
proposed mechanism is therefore a two-stage evaporation process: first from
the SSD into the warm layer, and then from the warm layer (at smaller radii)
into the \ISAF{} (see Fig.~\ref{fig-geometry} for a sketch).

In the paper of Spruit \& Deufel `zero dimensional' estimates were given for
the three-component flow geometry, without actual computation of the flow
geometry. This is the topic of the present paper. We solve the combined
continuity- and angular-momentum conservation equations for all three
components (\ISAF{}, warm flow and SSD) simultaneously with the appropriate
mass exchange terms between them. We derive stationary solutions to these
combined equations and study their properties.

\section{Basic model}
We treat all three components (\ISAF{}, warm flow and SSD) in a vertically
integrated way, but we allow the components to coexist with each other at
every radius. Therefore, each component is represented by a surface density
function $\Sigma_{\mathrm{x}}(R,t)$ where $R$ is distance from the center,
$t$ is time and x stands for either ``\ISAF{}'', ``warm flow'' or
``SSD''. Both the warm flow and the SSD may be truncated at an inner radius
(larger than the last stable orbit). These radii, called $\rwarmin$ for the
warm layer and $\rssdin$ for the SSD may vary with time. In the stationary
state they are eigenvalues of the problem and depend on the global accretion
rate.  The temperature of the \ISAF{} is assumed to be a constant fraction
times the local virial temperature:
\begin{equation}
\tisaf = 1.26 \frac{\sqrt{1+0.56\,\alpha^2}-1}{\alpha^2} \frac{G M}{R}\frac{\mu
m_p}{k}\comma
\end{equation}
which can be derived from self-similar solutions of ADAFs using $\gamma=1.5$
(Narayan \& Yi \citeyear{narayanyi:1994}). The temperature of the SSD is
calculated self-consistently assuming Kramer's opacity:
\begin{equation}
T_{\ssd}(R) = 1.17\times 10^{8} \left( \alpha \sigssd^3 \frac{1}{\sqrt{\mu}}
\frac{M/M_{\odot}}{R^3}\right)^{1/7}
\fullstop
\end{equation}
The temperature of the warm layer is assumed to be 300 keV for $R<\rssdin$
and 60 keV for $R\ge \rssdin$. These values are estimates based on the
results from Deufel et al.~(\citeyear{deufeldulspruit:2002}), who calculated
the detailed vertical structure of the warm flow.

\subsection{Evaporation of warm layer into \ISAF{}}
Spruit \& Deufel (\citeyear{spruitdeufel:2002}) derived the evaporation rate
of the warm layer inward of $\rssdin$ into the \ISAF{}. They find the
following expression (valid for $R<\rssdin$):
\begin{equation}
\evapwarmadaf = \frac{m_e^{3/2}k^2}{8 m_e^{1/2}e^4\ln\Lambda}
\alpha^2 \Omega_K^2 T_e^2
\comma
\end{equation}
(where $kT_e\approx 300$ keV). The evaporation rate is therefore constant
between the inner radius of the warm layer and the inner radius of the cool
disk. The inner radius of the warm layer $\rwarmin$ may be located at the
last stable orbit at $R=6 GM/c^2$, but it can also be larger if the
evaporation process was so effective that it evaporates the layer before it
has reached the black hole. At this stage, we assume the latter, i.e.~that the
warm layer has $\rwarmin>6 GM/c^2$. The consistency of this assumption will be
part of the results obtained. 

\subsection{Evaporation of SSD into warm layer}
The evaporation from the SSD into the warm layer for $R>\rssdin$ is a more
subtle matter. This evaporation is powered by the ion bombardment of the SSD
by the \ISAF{}. The flux of ions from the \ISAF{} onto the SSD (and through
the warm layer) is:
\begin{equation}
\fion = \xi \rhoisaf \sqrt{\frac{k \tisaf}{\mu m_p}}
\comma
\end{equation}
which is defined as the sum of the fluxes on both sides of the disk. The
parameter $\xi$ is a dimensionless efficiency parameter of order unity. If
$\xi=1$, the \ISAF{} loses its matter at the fastest possible rate: any
proton that reaches the disk surface is lost. Effects such as tangled
magnetic fields or proton-proton collisions may reduce the loss rate of
matter to the disk, which is simulated by taking $\xi<1$.

The protons that enter the SSD, leave their energy in the disk's upper
layers. A fraction $\eta$ of this energy is radiated away by Bremsstrahlung
and possibly other processes. It is not straightforward to compute this
value $\eta$, so we keep it a parameter. The remainder of this energy can be
used for the heating of disk matter up to the temperature of the warm
layer. This creates an evaporation rate $\psi_{\ssd\rightarrow\warm}$ of:
\begin{equation}
\psi^0_{\ssd\rightarrow\warm} = (1-\eta) \psi_{\isaf\rightarrow\ssd}
\frac{T_{\isaf}-T_{\warm}}{T_{\warm}-T_{\ssd}} 
\fullstop
\end{equation}

From the static solutions of Deufel et al.~(\citeyear{deufeldulspruit:2002})
we know that, as $\Sigma$ approaches a value close to unity (let us define
this value as $\Sigma_0$), the evaporation will cease and a static solution
sets in. In this static solution the energy input by ion bombardment is
entirely compensated by radiative losses (i.e.~radiative losses in excess of
the $\eta$ fraction already taken into account in the above evaporation rate
formula). We simulate this saturation effect by multiplying
$\psi_{\ssd\rightarrow\ssd}$ by a factor $(1-\Sigma_{\warm}/\Sigma_0)$. The
mass gain of the warm layer is then:
\begin{equation}
\psi_{\ssd\rightarrow\warm} = \psi^0_{\ssd\rightarrow\warm}
\left(1-\frac{\Sigma_{\warm}}{\Sigma_0}\right)
\fullstop
\end{equation}
Here, and in the remainder of this paper, we take $\Sigma_0=6$ g/cm$^2$
(accounting for two sides of the Shakura-Sunyaev disk, each having a layer
of maximum surface density of $3$ g/cm$^2$.).

\subsection{Equations for all three components}
All three layers (disk, warm layer and \ISAF{}) obey their own equations of
continuity and angular momentum conservation, with the appropriate sources
terms to account for the exchange of matter between the them. The angular
momentum exchange is given by the mass exchanged and the difference in specific
angular momentum; viscous coupling between the layers is neglected. The
continuity equation is best expressed as an equation for the local mass
flux $\dot M_\xx(R)$ (where $\xx$ stands for either `$\ssd$',
`$\warm$' or `$\isaf$'). The equation for this is:
\begin{equation}\label{eq-mdot-integral}
\dot M_{\xx}(R) = \dot M_{\xx}(R_{0\mathrm{x}}) -
2\pi \int_{R_{0\mathrm{x}}}^{R} \psi_{\xx}(R) R dR
\comma
\end{equation}
where $R_{0\mathrm{x}}$ is the inner edge of phase $\xx$, which is $R_{\in}$
for the \ISAF{}, $R_{\rim}$ for the warm layer and $R_{\ssd}$ for the
accretion disk. The quantity $\psi_{\xx}$ is the mass source term. For the
SSD one has $\psi_{\mathrm{ssd}}\equiv \fion-\evapssdwarm$, for the warm
layer one has $\psi_{\mathrm{warm}}\equiv \evapssdwarm-\evapwarmadaf$ and
for the \ISAF{} one has $\psi_{\mathrm{isaf}}\equiv \evapwarmadaf-\fion$.

The integration constant $\dot M_{\xx}(R_{0\xx})$ is zero for the warm layer and
the SSD, since both have clearly defined inner edges due to evaporation. But
it is non-zero for the \ISAF{}, since this is the layer that in the end
carries all the matter into the black hole. For the \ISAF{} the value of
$\dot M_{\isaf}(R_{\in})$ is determined as an eigenvalue of the problem: by
demanding a vanishing mass flux at the outer edge $\dot
M_{\isaf}(R_{\out})=0$, the value of $\dot M_{\isaf}(R_{\in})$ is indirectly
determined. 

In the thin disk approximation the equation for angular momentum conservation
with a no-friction boundary condition at $R=R_{0\xx}$ can be cast into the
general form:
\begin{equation}\label{eq-signu-integral}
\Sigma_{\xx}\nu_{\xx} = \frac{1}{6\pi\sqrt{R}}\int_{R_{0\xx}}^{R} \frac{\dot
M_{\xx}(R)}{\sqrt{R}} dR 
\comma
\end{equation}
In order to solve the $\Sigma_{\xx}$ from this, one must evaluate the
viscosity coefficient $\nu_{\xx}$, which for a standard $\alpha-$disk is
given by
\begin{equation}
\nu_{\xx}  = \alpha \sqrt{\frac{R^3}{GM}} 
\frac{kT_{\xx}}{\mu m_p}
\fullstop
\end{equation}

\subsection{Numerical approach}
The above equations are solved in a set of nested subroutines and
functions. At the deepest level we solve the structure of the \ISAF{} for a
given evaporation rate $\psi_{\warm\rightarrow\isaf}(R)$ and given $\dot
M_{\isaf}(R_{\in})$. This function returns the value of the \ISAF{} mass
flux at infinity $\dot M_{\isaf}(R\rightarrow\infty)$. By feeding this
function into a root-finding routine (we use {\tt zbrent} from Numerical
Recipes) the value of $\dot M_{\isaf}(R_{\in})$ is found for which $\dot
M_{\isaf}(R\rightarrow\infty)=0$. Note that this value is, due to
recondensation of the outflowing part of the \ISAF{} material, smaller than
the total evaporation rate $2\pi\int \psi_{\warm\rightarrow\isaf}(R)
RdR$. In other words: some material goes more than once through the process
of evaporation. In any case, the above procedure solves the \ISAF{}
structure for given $\psi_{\warm\rightarrow\isaf}(R)$

One step higher in the hierarchy we solve the structure of the warm layer,
for given values of $R_{\warm}$ and $R_{\ssd}$. First we integrate the
equations from $R_{\warm}$ to $R_{\ssd}$ and determine
$\psi_{\warm\rightarrow\isaf}(R)$. We then solve for the structure of the
\ISAF{} in the way described above. The \ISAF{} spreads outward beyond
$R_{\ssd}$ and injects its protons into the SSD. This proton injection
provides the energy needed to evaporate the SSD into the warm layer, and
therefore determines the evaporation rate $\psi_{\ssd\rightarrow\warm}$.
This then enables us to continue the integration of the equations for the
warm layer from $\rssdin$ out to $R\rightarrow\infty$.

The function that computes the structure of the warm layer in the above way
returns the value of $\dot M_{\warm}(R\rightarrow\infty)$. We feed this
function into {\tt zbrent} so that the value of $R_{\warm}$ for which $\dot
M_{\warm}(R\rightarrow\infty)=0$ is automatically determined.  Note that
this is a nested use of {\tt zbrent}, since within the function that
determines the structure of the warm layer we also call {\tt zbrent} for
determining the structure of the \ISAF{}.

\section{Resulting solutions}\label{sec-result-struct}
\begin{figure}
\centerline{
\includegraphics[width=9cm]{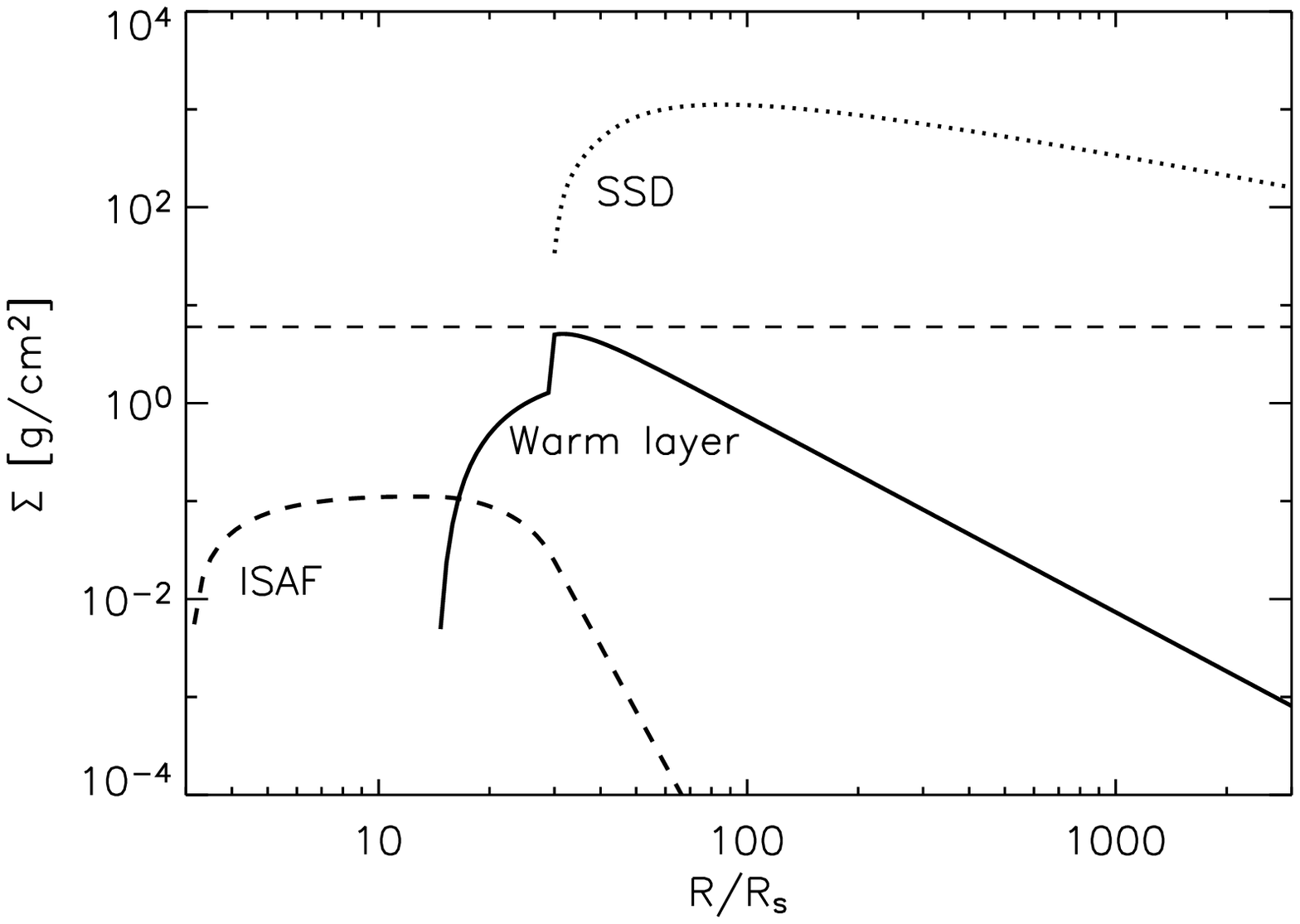}}
\centerline{
\includegraphics[width=9cm]{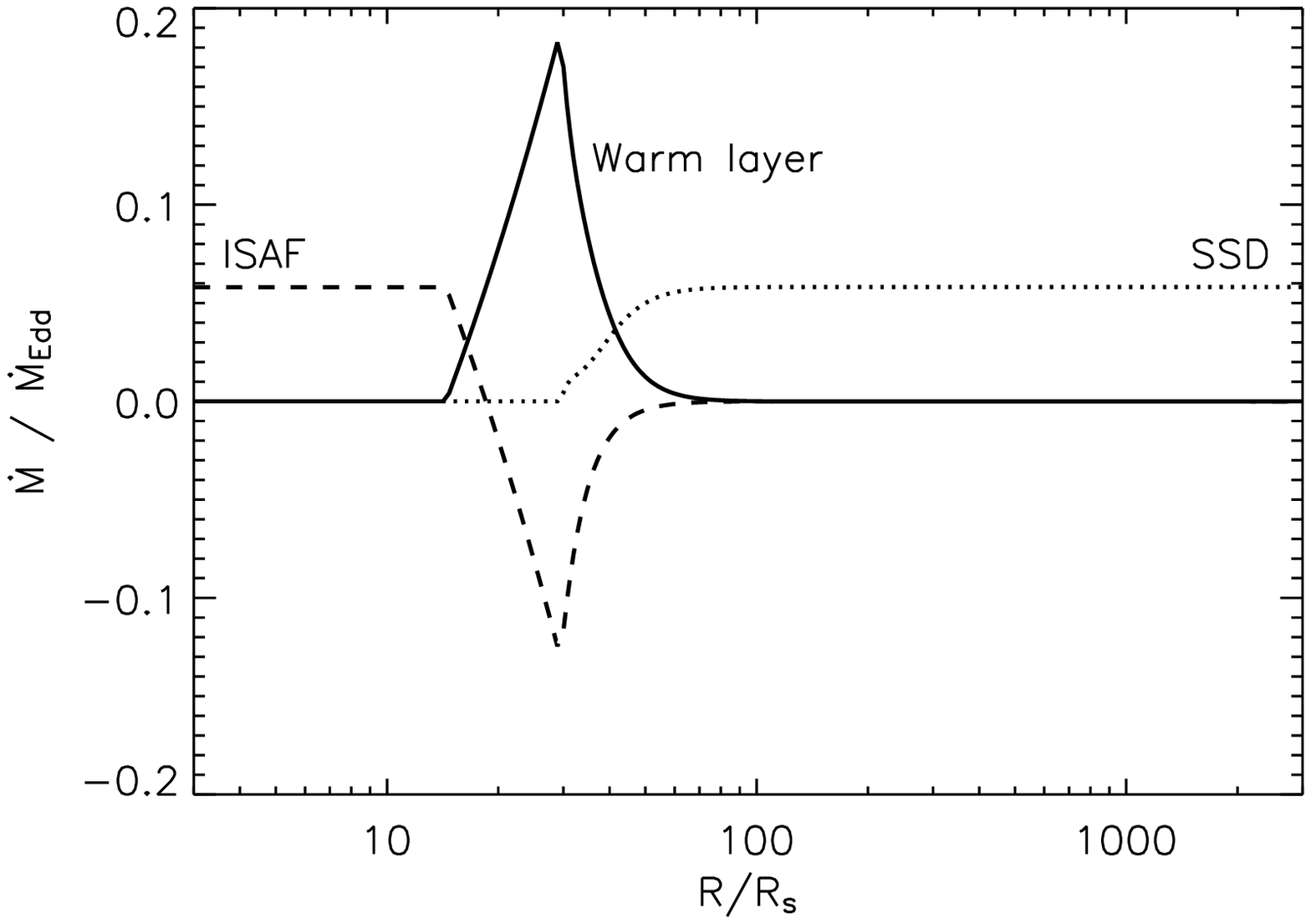}}
\caption{\label{fig-example-1}The radial three-layered structure of the
evaporating disk and the hot accretion flows, for the case of low $\alpha$
\revised{(i.e.~$\alpha=0.05$)} for the warm layer and the \ISAF{} (model
1). Upper panel shows the surface density of the two components: the cool
(SSD) disk, warm surface layer and the \ISAF{}. Lower panel shows the local
mass flux of the components in units of the Eddington rate (the Eddington
accretion rate is defined here using an efficiency factor of 0.1). The
horizontal dashed line in the top figure is the value of $\Sigma_0=6$
g/cm$^2$.}
\end{figure}

\begin{figure}
\centerline{
\includegraphics[width=9cm]{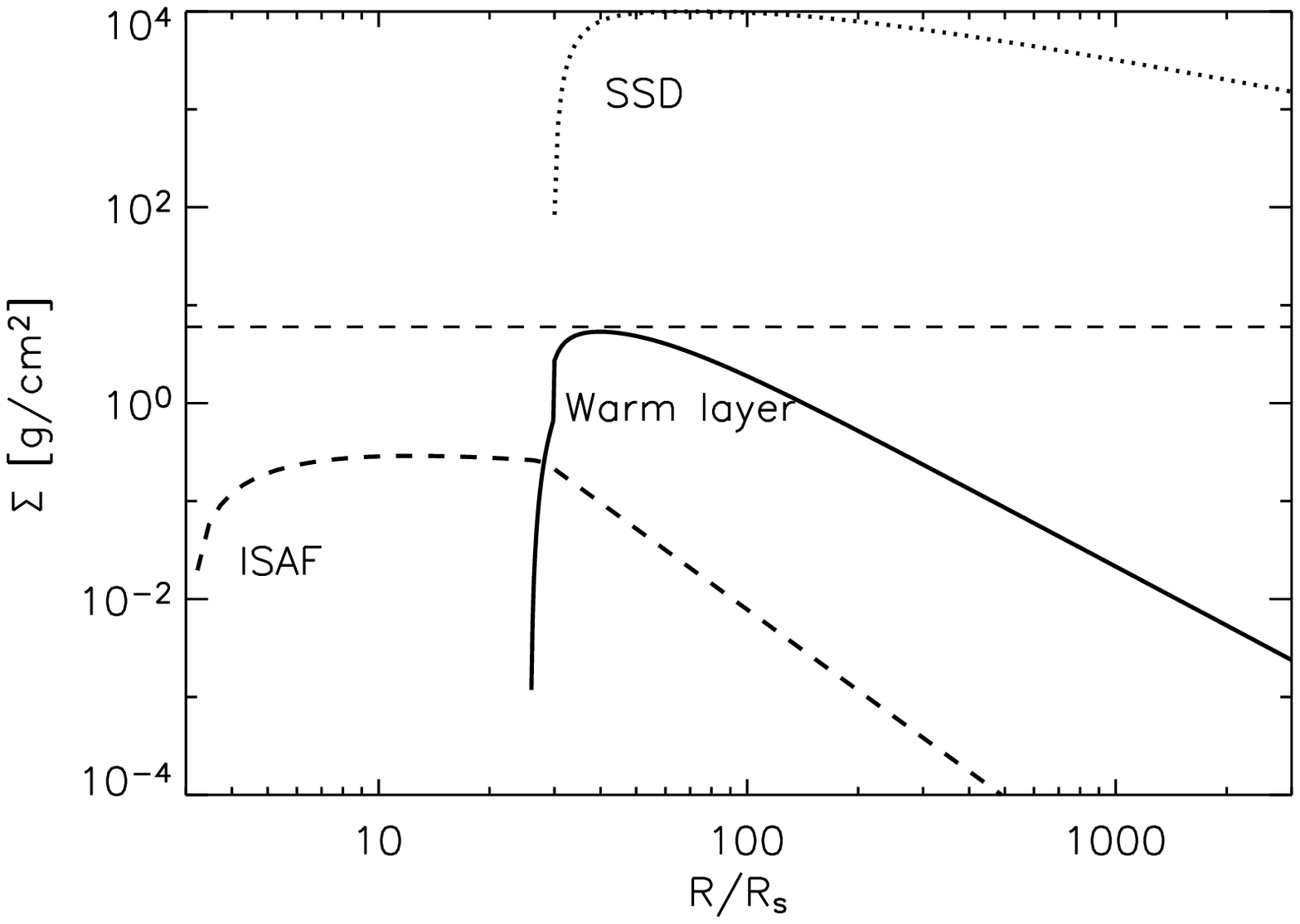}}
\centerline{
\includegraphics[width=9cm]{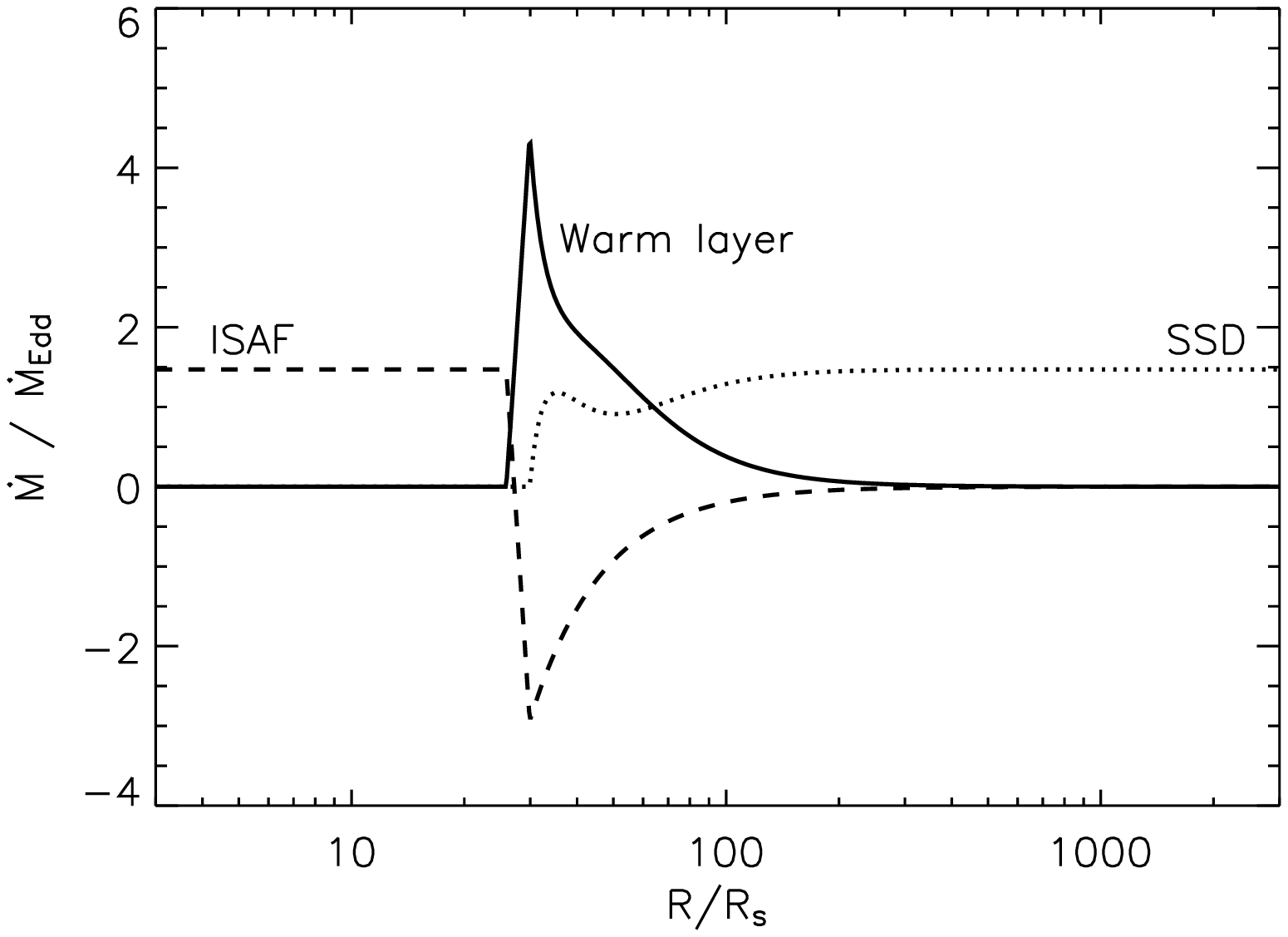}}
\caption{\label{fig-example-2}The radial structure for the case of
$\alpha=0.5$ (model 2).}
\end{figure}

\begin{figure}
\centerline{
\includegraphics[width=9cm]{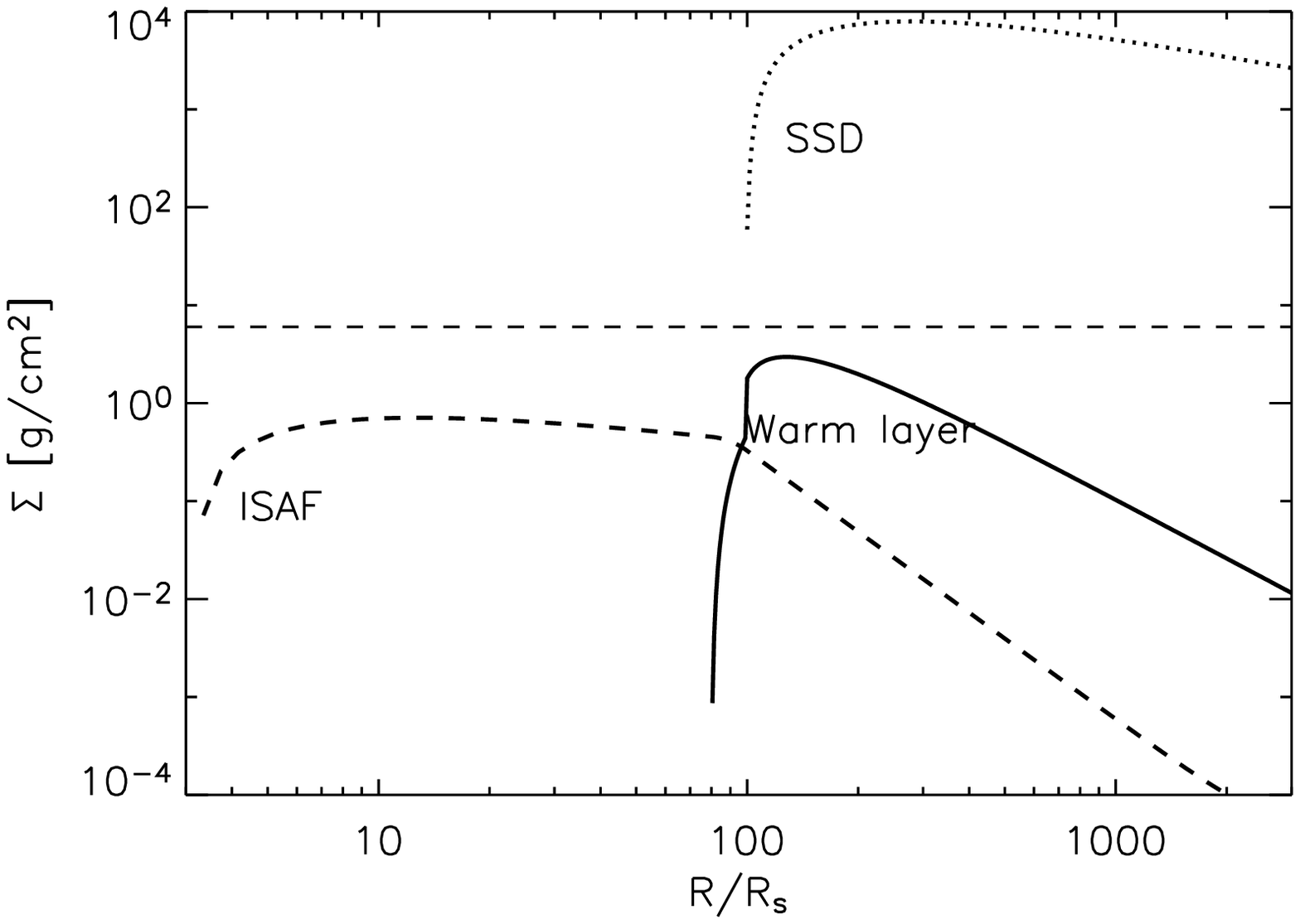}}
\centerline{
\includegraphics[width=9cm]{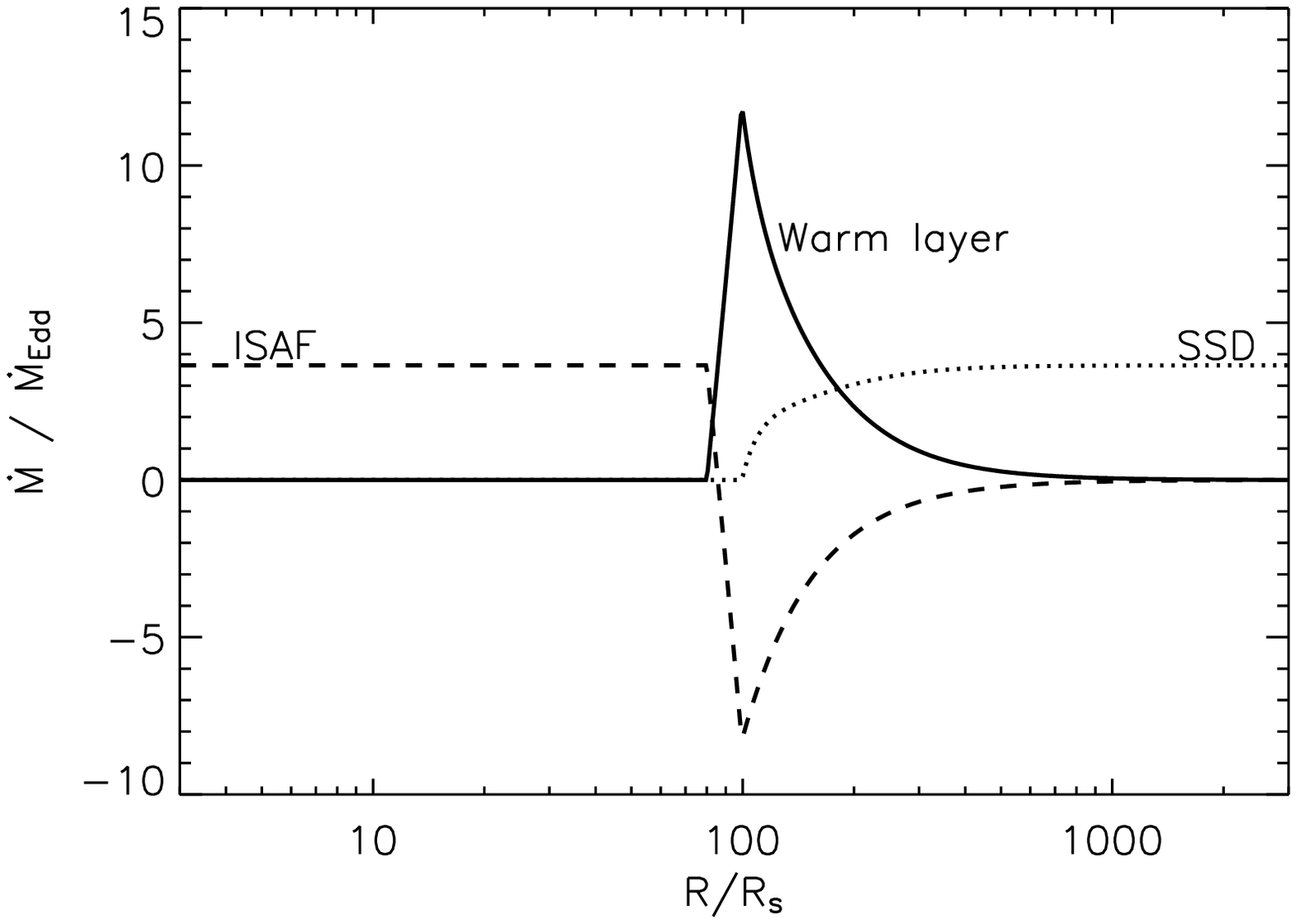}}
\caption{\label{fig-example-3}\revised{The radial structure for the case of
$\alpha=0.5$ (like model 2) but this time with $R_{\ssd}=100\,R_s$ (model
3).}}
\end{figure}

Using the above procedure we find stationary solutions for most (though not
all) parameter combinations. In Figs.~\ref{fig-example-1},
\ref{fig-example-2},\revised{\ref{fig-example-3}} we show \revised{three}
examples, \revised{all} of which have $M_{\mathrm{bh}}=10\,M_{\odot}$,
$\xi=1$ and $\eta=0.9$. In \revised{model} 1 (Fig.~\ref{fig-example-1}) we
take $\alpha=0.05$ for all three components, while in \revised{model} 2
(Fig.~\ref{fig-example-2}) the \ISAF{} and the warm layer have a ten times
higher viscosity: $\alpha=0.5$. \revised{In both these models we take
$R_{\ssd}=30\,R_s$. In model 3 we take $\alpha=0.5$ and
$R_{\ssd}=100\,R_s$.}  As can be seen in these figures, the mass flux $\dot
M$ for the \ISAF{} is positive inward of a certain radius, and negative
outward of that radius. This negative mass flux is the outflow in the
\ISAF{} driven by the angular momentum conservation. It rapidly goes back to
zero toward larger radii due to the loss of \ISAF{} ions into the SSD. The
mass flux of the warm layer is zero at both small and large radii, and has a
peak value at the inner edge of the SSD ($\rssdin$). Beyond $\rssdin$ the
mass of the SSD is evaporated into the warm layer, while at $R<\rssdin$ the
warm layer evaporates into the \ISAF{}. The surface density of the warm
layer makes a jump at $\rssdin$, which is due to the temperature jump (300
KeV inward of $\rssdin$ and 80 KeV outward of $\rssdin$).
In the case of low viscosity (\revised{model 1,} Fig.~\ref{fig-example-1})
the warm layer only narrowly reaches the saturation surface density
$\Sigma_{\mathrm{warm}}=\Sigma_0\equiv 6.0$, but for the high viscosity case
(\revised{model 2,} Fig.~\ref{fig-example-2}) this saturation value is
reached easily. \revised{For the same high $\alpha$ but with larger
$R_{\ssd}$, the surface density does not reach saturation anymore.}

Interestingly, this saturation of $\Sigma_{\mathrm{warm}}$ causes the mass
flux in the SSD to acquire an peculiar wiggle \revised{in model 2} (clearly
seen in Fig.~\ref{fig-example-2}): first (at large radii) the matter in the
SSD is evaporated into the warm layer ($d\dot M_{\ssd}/dR>0$), but as one
gets closer to the inner radius $\rssdin$, some matter of the warm layer is
dumped back onto the SSD ($d\dot M_{\ssd}/dR<0$).  Only just beyond
$\rssdin$ the evaporation of the SSD kicks in again and produces the cut-off
in $\dot M_{\mathrm{ssd}}$ and the maximum in $\dot M_{\mathrm{warm}}$ at
$\rssdin$. This recondensation takes place because if the surface density of
the warm layer has reached the saturation value $\Sigma_0$ (which is
independent of radius), the mass flux in the warm layer should go roughly
proportional to $R^{3/2}$ because both $\Sigma_{\mathrm{warm}}$ and the
temperature of the warm layer are then constant with $R$. A decreasing mass
flux in the warm layer in the flow direction ($d\dot M_{\warm}/dR<0$)
inevitably means recondensation onto the SSD.  This wiggle is not present in
the case of low viscosity or the case with large $R_{\ssd}$, since the
saturation value $\Sigma_0$ is not reached by the warm layer.

Another interesting difference between the example models is that the 300
KeV part of the warm layer (i.e.\ the part inward of $\rssdin$) is narrower
for the case of high viscosity (\revised{model 1}) than for the case of low
viscosity (\revised{model 2}).

\subsection{Global trends}
\begin{figure}
\centerline{
\includegraphics[width=9cm]{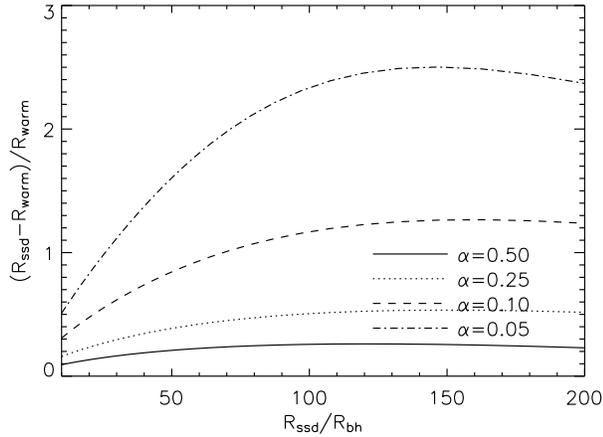}}
\caption{\label{fig-dr-afo-rssd}The width of the 300 KeV region of the
warm layer as a function of $\rssdin$. This width is defined as 
$\rssdin/\rwarmin-1$.}
\end{figure}

The dependence of some global parameters of the models on the input
parameter $\rssdin$ are discussed in this subsection. In
Fig.~\ref{fig-dr-afo-rssd} the width of the 300 KeV part of the warm layer
(i.e.~the part inward of $\rssdin$) is shown as a function of $\rssdin$. It is
clear that for low $\alpha$ and for large $\rssdin$ this width is the
largest. Since this region may be responsible for the production of hard
X-rays, its surface area is an important quantity of the model. Also, if
this region is wide, then the approximation that the soft photons of the SSD
do not cool this region (hence allowing it to have 300 KeV temperature) is
more valid than in the very narrow case.

\begin{figure}
\centerline{
\includegraphics[width=9cm]{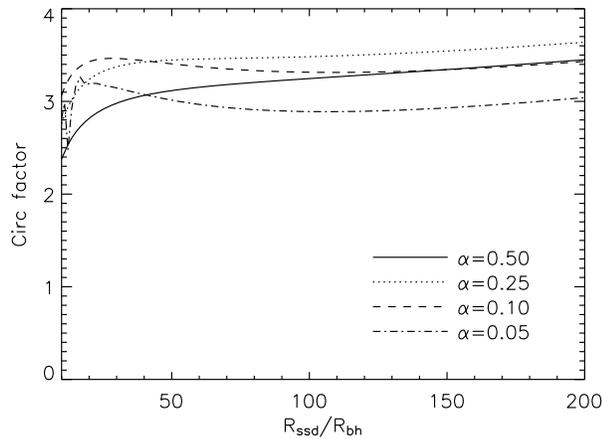}}
\caption{\label{fig-circ-afo-rssd}The circulation rate within the stationary
solution.}
\end{figure}

In Fig.~\ref{fig-circ-afo-rssd} the circulation rate is shown. It tells how
often the matter is evaporated and recondensed again 
before accreting onto the hole. As one can see, for
all values of $\alpha$ and $\rssdin$ this rate is always of the order of 3,
meaning that on average the \ISAF{} matter gets injected into the SSD (and
re-evaporated from the warm layer) twice before finally flowing into the
black hole. In other words: 2/3 of the hot \ISAF{} matter is used up for the
proton injection into the SSD. 

\subsection{Radiative output}
\begin{figure}
\centerline{
\includegraphics[width=9cm]{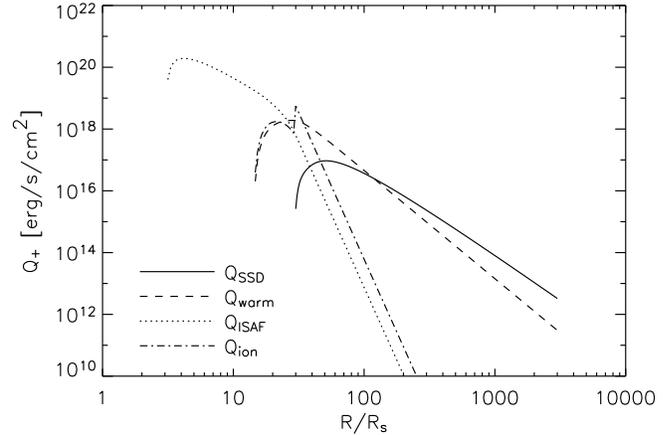}}
\caption{\label{rads} Energetics of the three components of the accretion
flow for model 1 (i.e.~$\alpha=0.05$ for all components).  Solid: radiation
from the cool disk, dashed: viscous dissipation in the warm component,
dot-dashed: heating of the warm layer by ion illumination from the \ISAF,
dotted: viscous dissipation in the \ISAF.  Since the \ISAF{} is radiatively
inefficient, the total radiative output from the flow is dominated by the
hard radiation from the warm layer.  The accretion rate for this case is
$\dot m= 0.06$ (in units of Eddington).}
\end{figure}

The three components of the flow each contribute to the X-ray spectrum.  The
cool disk produces a very soft component, the warm layer and the \ISAF{} a
hard component (peaking above 100 keV). A detailed calculation of the
overall spectrum produced by the combination of these components would be
somewhat premature, in view of the uncertainties in the model. The relative
contributions of hard and soft X-rays, however, can be determined simply
from the energy budgets of these components. Fig.~\ref{rads} shows these in
the form of the vertically integrated heating rates of the components, per
unit surface area. The \ISAF{} has the largest amplitude, since the energy
release in it takes place in the deepest part of the gravitational
potential. The radiative efficiency of an \ISAF{} is low, however, so that
the radiative output is actually dominated by the other two components of
the flow, in particular the warm component. The exact value of the radiative
efficiency of an \ISAF{} depends on details, but typical numbers are of the
order $\epsilon\approx \dot m\alpha^2$ (e.g.~Rees et
al.~\citeyear{reesbegblph:1982}; Esin et
al.~\citeyear{esinmcclnar:1997}). For the case shown in Fig.\ \ref{rads},
this yields $\epsilon\approx 1.5\times 10^{-4}$. The contribution of the
\ISAF{} is thus small, and most hard radiation comes from the warm
layer. The total energy deposition into the warm layer (viscous heating plus
ion illumination) is $Q_{\mathrm{tot,warm}}=1.2\times 10^{35}$ erg/s for the
example of Fig.\ \ref{rads} (model 1). All of this goes into hard radiation,
but a fraction of this radiation is intercepted by the cool disk and
reprocessed into soft flux. The internal viscous dissipation in the cool
disk is a smaller contribution ($Q_{\mathrm{tot,ssd}}=4.6 \times 10^{34}$
erg/s). In this way, we find, for the example of Fig.\ \ref{rads}, $F_{\rm
hard}/F_{\rm soft}\approx 2.6$.
The cool disk produces a standard multi-color disk spectrum; since its inner
edge is at some distance from the hole, its contribution to the X-ray
spectrum is very soft. Most of the energy dissipated in the warm layer is
converted to to hard X-rays by Comptonization of soft photons from the cool
disk (a fraction is spent in poducing soft flux by irradiation of the cool
disk surface). The \ISAF{} also produces a hard spectrum, but since it is
radiatively inefficient, its contribution to the total hard flux is much
smaller than its energy budget as shown in Fig.~\ref{rads}. 

The conclusion is thus that the model produces a predominantly hard
spectrum. The soft flux produced is mostly due to reprocessing of hard flux
by the cool disk directly underneath the warm layer, but includes a small
contribution from internal viscous heating in the disk. The hard flux comes
mostly from the region where the warm layer evaporates into the \ISAF,
i.e. a ring just inside the inner edge of the cool disk. This is consistent
with the close connection found observationally between the timing
properties of the hard flux and the inner radius of the cool disk
(e.g. Gilfanov et al.~\citeyear{gilfchurrev:2000}; Churazov et
al.~\citeyear{churgilfrev:2001}; Revnivtsev et
al.\citeyear{revgilfchur:2001}).

\subsection{Stability}

\subsubsection{Stability of the vertical structure}
If we envisage the radial structure of the cool disk as fixed, one can study
whether the structures of the warm and hot (\ISAF{}) layers is stable to
small perturbations. This question was answered in part by Deufel et
al.~(\citeyear{deufeldulspruit:2002}). In these calculations, the vertical
structure of a cool disk plus warm layer under ion illumination was
determined in a time dependent way, but without the radial structure that we
have included here. In these calculations, the results quickly converge to a
stable static equilibrium, as long as the cool disk is present. The
adjustment takes place on the thermal time scale of the warm layer. This
stability can be understood as a consequence of the overall energy balance
(Haardt et al.~\citeyear{haardtmarghis:1994}) between comptonization in the
warm layer and soft photon production in the cool disk. Inside the inner
edge of the cool disk, however, the warm layer becomes unstable due to the
lack of soft photon flux, and transforms into the evaporating flow studied
here.

\subsubsection{Stability of the radial structure}
In the present calculations, additional questions of stability arise related
to the radial structure of the accretion flows. The first question we posed
is if the combination of \ISAF, warm layer and cool disk is stable if the
position of the inner edge of the cool disk is kept fixed. In other words,
if the evaporation into the \ISAF, its spreading over the cool disk and
condensation onto it converges to a stable stationary state. Preliminary
time-dependent simulations of the warm disk, for a fixed accretion disk, and
for a stationary \ISAF{} computed at every time-step, show that the warm
layer seems to be stable, and will find the stationary solution quickly. But
full time dependent simulations of both hot phases (the warm disk and the
\ISAF{}) have not been carried out to confirm this.

\begin{figure}
\centerline{
\includegraphics[width=9cm]{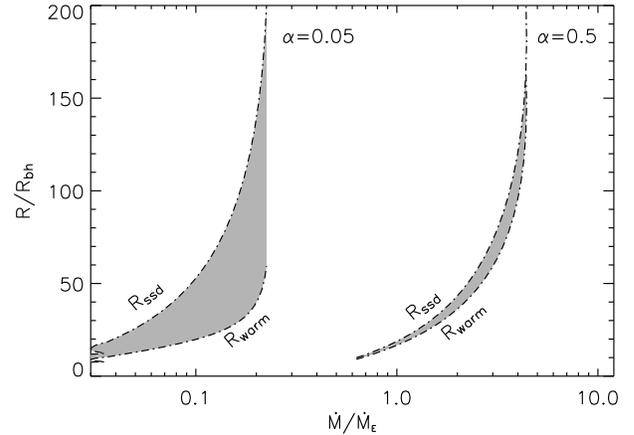}
}
\caption{\label{fig-mdot-afo-rssd}
\revised{The inner radius of the cool disk $\rssdin$ and the inner radius of
the warm phase $\rwarmin$ as a function of global accretion rate $\dot M$ in
units of the Eddington accretion rate $\dot M_E$ (defined using a 10\%
efficiency factor). The grey area shows therefore the extend of the
300 KeV evaporating inner part of the warm layer inward of the cool
disk. Shown are the results for models 1 ($\alpha=0.05$) and 2 
($\alpha=0.5$). Note that in the modeling procedure the $\rssdin$ is
the model parameter and $\dot M$ the result. But since in nature it is
the opposite, we plot $\dot M$ on the x-axis here.}
}
\end{figure}

A second question is whether the position of the inner edge of the cool disk
is stable: on time scales longer than the adjustment time of the warm layer,
the balance between evaporation and condensation may lead to a drift of the
position of the inner edge of the cool disk. This question can be answered
already without a full time dependent study by considering the relation
between the radius of the inner edge of the accretion disk and the total
mass accretion rate. This is shown in Fig.~\ref{fig-mdot-afo-rssd}.
\revised{It shows that for increasing accretion rate the inner radius of the
cool disk (as well as the inner radius of the warm layer) increases as
well. The reason is that in our model the warm flow near $\rssdin$ must
carry the entire mass accretion rate $\dot M$ while it's surface density is
limited to at most $\Sigma_0$. For higher $\dot M$ it therefore requires a
larger circumference to transfer all the matter with the same surface
density, hence the larger $\rssdin$.}

\revised{This relation is opposite to what is found from the coronal
evaporation mechanism of Meyer \& Meyer-Hofmeister (1994), where the inner
radius of the cool disk decreases for increasing global accretion rate.}
This relation caused the Meyer \& Meyer-Hofmeister evaporation mechanism to
be self-regulating: if the $R_{\ssd}$ is too large, it will decrease until a
balance is reached between the total mass accretion rate in the disk and the
evaporation rate.  Such a self-regulatory mechanism is not present in
ion-bombardment evaporation process. A small deviation from the stationary
solution may let the $\rssdin$ move e.g.~slightly inward, which decreases
the allowed global accretion rate for a stationary solution. Since the
accretion rate is fixed from the outside and cannot be changed, it will
exceed this new value and therefore decrease $\rssdin$ even further. We
therefore expect that a small deviation of $R_{\ssd}$ from the equilibrium
value would amplify and cause the inner edge of the SSD to move inwards or
outwards in a run-away fashion. The time scale to be expected for this
instability is the viscous time scale of the SSD.

The question thus arises whether the evaporating solutions found here can
survive in practice; in any case it is clear that some additional piece of
physics will be needed. This need not be seen as a debilitating
disadvantage, however, since the variability of the hard state shows that
the accretion flow is in fact violently unstable. If the evaporation
mechanism had predicted a stable stationary flow, it would have needed
additional physics as well, in this case something to explain the observed
instability of the flow.

\section{Discussion and conclusions}
In this paper we have presented the structure of the 3-layered accretion
disk model envisaged by Spruit \& Deufel (\citeyear{spruitdeufel:2002}).
This models is based on the ion-illumination model for X-ray binaries in
which a hot ion supported flow coexists with a cool accretion disk (Deufel
et al.~\citeyear{deufeldulspruit:2002}; Spruit and Deufel
\citeyear{spruitdeufel:2002}). This coexistence leads to a three-layered
structure: between the cool disk and the ion supported flow a `warm' layer
is formed by ions penetrating the disk from the ion supported flow. The ion
and electron and electron temperatures are in equilibrium in this warm layer
(in contrast with ion supported flow, where the ion temperature is much
higher than the lectron temperatue). Its temperature, in the 100 keV range,
is governed by the equilibrium between ion heating and Compton cooling by
soft photons from the cool disk. This equilibrium is quite stable, in
contrast with a single optically thin accretion flows in this temperature
range, which are strongly thermally unstable (the well known instability of
the `SLE' branch of accretion solutions, e.g.~Chen et
al.~\citeyear{chenabrlas:1995}). In the paper of Deufel et
al.~(\citeyear{deufeldulspruit:2002}) the local vertical structure (for a
given distance from the hole) of the accretion flow was calculated. The
calculations presented here are complementary to these. They focus instead
on the radial structure of the accretion flow, while the vertical structure
is approximated by a 3-layer approximation. The exchange of energy and
angular momentum between these layers is taken into account, as well as the
processes of evaporation and condensation between the ion supported flow and
the warm layer, and between the warm layer and the cool disk.

We find that stationary solutions exist for a wide range of parameters, and
that these flow geometries exhibit circulation patterns in which matter is
evaporated and recondensed several times before vanishing into the black
hole. This finding is of special significance for the Li-enhancements
observed in the secondaries of black hole binaries (Mart\'{\i}n et
al.~\citeyear{martinrebolo:1994}), which can be explained as a result of
Li-production in the accretion disk by ion illumination (spallation of CNO
nuclei and He+He fusion), provided that in addition to the accretion flow
there is an outflow from the inner regions of the accretion disk that can
carry the Li to the secondary. This is an attractive possibility since the
hard spectral states that we associate here with the ion-illumination
process is known to be closely related to the presence of jets in in black
hole binaries. The `recirculation' of matter through the ion-bombarded warm
layer increases the Li-production significantly.

We find that the energy radiated by the ion-illuminated, evaporating flow is
likely to be dominated by a hard X-ray component produced in the evaporating
region itself. A soft component is also produced by reprocessing in the cool
disk, but it is energetically less important than the hard component. This
provides a satisfactory solution to a long-standing problem in models where
the hard and soft fluxes are coupled through reprocessing. The generic model
for this coupling (Haardt and Maraschi) has the property of predicting
similar energy fluxes in the hard and soft components. In order to make the
hard component dominate as observed, modifications of the geometry are
needed to reduce the interaction between the soft photons and the
Comptonizing region. In the evaporating flows presented here, this occurs
naturally because most of the hard flux is produced in a `soft photon
staved' region just inside the inner edge of the cool disk.

We also show that the globally stationary solutions are not stable on long
time scales. Eventually the inner rim of the evaporating SSD will move
either outward or inward in a run-away fashion. However, if the inner rim of
the SSD is kept fixed, the structures of the warm intermediate layer and the
hot \ISAF{} are presumably stable. Hence, this evaporation mechanism is
feasible, but it will result in time-dependent phenomena instead of a
stationary state. How these time-dependent solutions behave remains the
topic of a future detailed time-dependent study of ion-illuminated
evaporating disks.

\end{document}